\documentclass[aps,twocolumn,showpacs,superscriptaddress,amsfonts,amssymb,amsmath]{revtex4}

\begin{document}

\preprint{Yukawa Institute Kyoto}
\preprint{YITP-06-63}

\title{Deformed Fokker-Planck Equations}

\author{Choon-Lin Ho}
 \affiliation{Department of Physics, Tamkang University,
 Tamsui 251, Taiwan, Republic of China}
\author{Ryu  Sasaki}
 \affiliation{Yukawa Institute for Theoretical Physics,
     Kyoto University, Kyoto 606-8502, Japan}

\date{Dec 13, 2006}

\begin{abstract}

 Based on the well-known relation between
Fokker-Planck   equations and Schr\"odinger equations
of quantum mechanics (QM), we propose new {\em deformed \/}
Fokker-Planck (FP) equations associated
with the Schr\"odinger equations
of ``discrete" QM.
The latter is a natural discretization of QM and its
Schr\"odinger equations are
difference instead of differential equations.
Exactly solvable FP equations are obtained corresponding
to exactly solvable ``discrete" QM,
whose eigenfunctions include various deformations of the classical
orthogonal polynomials.

\end{abstract}


 \maketitle

{\bf 1.}~Since it was first introduced by Fokker and Planck to describe
the Brownian motion of particles, the Fokker-Planck (FP) equation
has become one of the basic tools used to deal with fluctuations
in various kinds of systems \cite{FP}.  Recently, the phenomena of
anomalous diffusion in fractal and disordered media have prompted
new developments in the FP theory.  Most attempts have been along
developing fractional FP equations, where the ordinary spatial
derivatives are replaced by fractional derivatives \cite{FracFP}.
Others have tried to generalize the linear FP equation to
nonlinear ones \cite{NFP}.  Diffusion equation based on
$q$-derivatives has also been considered \cite{Ho}.

In this paper, we shall generalize the FP equation in a different
direction.  We shall derive new types of deformed FP equation
which are associated with the Schr\"odinger equations of the
``discrete" quantum mechanics considered in \cite{OS1,OS2}.
The eigenfunctions of some exactly solvable ``discrete" quantum mechanics
include various  deformations of the classical orthogonal polynomials
(the Hermite, Laguerre and Jacobi polynomials), namely,
those belonging to the family of Askey-scheme of hypergeometric
orthogonal polynomials, and the Askey-Wilson polynomial in
$q$-analysis \cite{AAR,KS,AW}.
The discrete Schr\"odinger
equation is a discrete deformation of the usual Schr\"odinger
equation in much the same way that the Ruijsenaars-Schneider-van
Diejen type systems \cite{RSD} are the ``discrete" counterparts of
the Calogero and Sutherland systems \cite{CS}, the celebrated
exactly solvable multi-particle dynamics.
In fact, these deformed orthogonal polynomials arise
in the problem of describing the equilibrium positions of
Ruijsenaars-Schneider  type
systems \cite{OS2}, corresponding to the facts that the Hermite, Laguerre
and Jacobi polynomials describe the equilibrium positions of the
Calogero and Sutherland systems \cite{zero}.
In this sense, our new
deformed FP equations can be considered as the ``discrete"
deformation of the usual FP equation.

\medskip

{\bf 2.}~In one dimension, the FP equation of the probability
density $P(x,t)$ is \cite{FP}
\begin{gather}
\frac{\partial P(x,t)}{\partial t}=L_{FP}P(x,t),\nonumber\\
L_{FP}\equiv -\frac{\partial}{\partial x} D^{(1)}(x) +
\frac{\partial^2}{\partial x^2}D^{(2)}(x).
 \label{FPE}
\end{gather}
The functions $D^{(1)}(x)$ and $D^{(2)}(x)$ in the FP operator
$L_{FP}$ are, respectively,  the drift and the diffusion
coefficient (we consider only time-independent case).  The drift
coefficient represents the external force acting on the particle,
while the diffusion coefficient accounts for the effect of
fluctuation.  The drift coefficient is usually expressed in terms
of a drift potential $\Phi(x)$ according to
$D^{(1)}(x)=-\Phi^\prime (x)$, where the prime denotes derivative
with respect to $x$.

It is well known that the FP equation is closely related to the
Schr\"odinger equation. The correspondence between the two
equations is usually given by transforming the FP equation into
the corresponding Schr\"odinger equation \cite{FP}. But the
argument could well be reversed.

Consider a quantum mechanical Hamiltonian of one degree of freedom
 (we adopt the unit system in which $\hbar$ and the mass $m$
of the particle are such that $\hbar=2m=1$)
\begin{equation}
H=-\frac{\partial^2}{\partial x^2}+V(x).
\end{equation}
Let $V(x)$ be such that the ground state energy is zero, i.e.
$H\phi_0=0$. The Hamiltonian $H$ is completely  determined by its
ground state wave function $\phi_0(x)$. By the well-known theorem
of quantum mechanics, $\phi_0(x)$ has no node and can be chosen
real. Thus it can be parametrized by a real {\em prepotential}
$W(x)$:
\begin{gather}
\phi_0(x)=e^{-W(x)},
\label{W}\\
V(x)=W^\prime (x)^2 - W^{\prime\prime}(x). \label{V}
\end{gather}
The Hamiltonian is factorizable and positive semi-definite: $H=A^\dagger A$.
Here $A^\dagger\equiv -\partial_x +W^\prime$ and $A\equiv \partial_x
+W^\prime$.
The constant part of $W(x)$ is so chosen as to normalize $\phi_0(x)$
properly, $\int\phi_0(x)^2\,dx=1$.

Now we define an operator from the $H$ and $\phi_0$ by the
similarity transformation
\begin{equation}
L_{FP}\equiv -\phi_0 H \phi_0^{-1}~\label{FP},
\end{equation}
which guarantees the non-positivity of the eigenvalues of
$L_{FP}$. With $\phi_0$ given in (\ref{W}), one obtains
\begin{equation}
L_{FP}=\frac{\partial}{\partial x}2W^\prime +
\frac{\partial^2}{\partial x^2}.
\end{equation}
From (\ref{FPE}), it is seen that $L_{FP}$ is just the
corresponding FP operator with $\Phi(x)=2W$ as the drift potential
and with a constant diffusion coefficient (here equals one owing
to the choice of unit in $H$). Hence both $H$ and $L_{FP}$ are
determined by $\phi_0$. The eigenfunction $P_n (x)$ of $L_{FP}$
corresponding to eigenvalue $-\lambda_n$ is related to the
(real and normalized) eigenfunction $\phi_n$ of $H$ corresponding to
$\lambda_n$ by
$P_n(x)=\phi_0(x)\phi_n(x)$.  The stationary distribution is
$P_0=\phi_0^2=\exp(-2W)$, which is obviously non-negative, and is
the zero mode of $L_{FP}$: $L_{FP}P_0=0$.  Any positive definite
initial probability density $P(x,0)$ can be expanded as
$P(x,0)=\phi_0(x)\sum_n c_n\phi_n(x)$, with constant coefficients
$c_n$ ($n=0,1,\ldots$)
\begin{eqnarray}
c_n=\int_{-\infty}^\infty \phi_n(x)\left(\phi_0^{-1}(x)
P(x,0)\right)dx.
\end{eqnarray}
Then at any later time $t$, the solution of the FP equation is
$P(x,t)=\phi_0(x)\sum_n c_n \phi_n(x)\exp(-\lambda_n t)$.

\medskip

{\bf 3.} We now derive a new class of deformed FP equation
corresponding to the discrete Schr\"odinger equations discussed in
\cite{OS1} in accordance with the prescription in the last
section.  Eigenfunctions in this type of discrete Schr\"odinger
equations are related to the family of Askey-scheme of
hypergeometric orthogonal polynomials \cite{AAR,KS}.

The Hamiltonian has the general form
\begin{eqnarray}
  H &\equiv& \sqrt{V(x)}\,e^{-i\partial_x}\sqrt{V^*(x)}
  +  \sqrt{V^*(x)}\,e^{i\partial_x}\sqrt{V(x)}\nonumber\\
  &&-\left(V(x)+V^*(x)\right).
  \label{H}
\end{eqnarray}
Here the momentum operator $p=-i{\partial_x}$ (with $\hbar=1$)
appears as exponentiated instead of powers in ordinary quantum
mechanics. Thus they cause a finite shift of the wave function in
the {\em imaginary\/} direction: $e^{\pm
i\partial_x}\phi(x)=\phi(x\pm i)$.  Throughout this paper the
following convention of a complex conjugate function will be
adopted: for an arbitrary function $f(x)=\sum_na_nx^n$,
$a_n\in\mathbb{C}$, we define $f^*(x)=\sum_n a_n^* x^n$.
Here $a_n^*$ is the complex conjugation of $a_n$.
Note that $f^*(x)$
is not the complex conjugation of $f(x)$, $(f(x))^*=f^*(x^*)$.
This is particularly important when a function is shifted in the
imaginary direction.

The Hamiltonian (\ref{H}) is factorised, i.e. $H=A^{\dagger}A$
with
\begin{gather}
 A\equiv e^{-\frac{i}{2}\partial_x}\sqrt{V^*(x)}
  -e^{\frac{i}{2}\partial_x}\sqrt{V(x)},\\
  A^{\dagger}\equiv \sqrt{V(x)}\,e^{-\frac{i}{2}\partial_x}
  -\sqrt{V^*(x)}\,e^{\frac{i}{2}\partial_x}.
\end{gather}
Here $\dagger$ denotes the ordinary hermitian conjugation with
respect to the ordinary $L^2$ inner product: $\langle
f|g\rangle=\int_{-\infty}^\infty(f(x))^*g(x)dx$. Obviously the
Hamiltonian (\ref{H}) is hermitian (self-conjugate) and positive
semi-definite.

Let the ground state $\phi_0$ be annihilated by $A$:
\begin{eqnarray}
  A\phi_0(x)=0,\quad (
  \Longrightarrow H\phi_0(x)&=&0,\quad \lambda_0=0).
  \label{phi0form}
\end{eqnarray}
Explicitly the above equation reads
\begin{equation}
  \!\sqrt{V^*\left(x-{i}/{2}\right)}~\phi_0\left(x-{i}/{2}\right)
  \!=\!\sqrt{V\left(x+{i}/{2}\right)}~\phi_0\left(x+{i}/{2}\right),
\label{Cond}
\end{equation}
or equivalently
\begin{eqnarray}
  \sqrt{V^*(x-i)}~\phi_0(x-i)&=&\sqrt{V(x)}~\phi_0(x),\label{Cond-1}\\
  \sqrt{V(x+i)}~\phi_0(x+i)&=&\sqrt{V^*(x)}~\phi_0(x).\label{Cond-2}
\end{eqnarray}
Eq.~(\ref{Cond}) relates the potential $V(x)$ and the ground state
$\phi_0$, and hence is the discrete analogue of (\ref{W}) albeit
in an implicit way.

Now, we form the associated FP operator from (\ref{H}) and
$\phi_0$ in (\ref{phi0form}) according to the similarity
transformation (\ref{FP}).  The sum $V(x)+ V^*(x)$ remains intact.
The first term in $H$ transforms as 
\begin{eqnarray}
&&\phi_0(x)
\sqrt{V(x)}\,e^{-i\partial_x}\sqrt{V^*(x)}\,\phi_0^{-1}(x)\nonumber\\
=&&\left(\frac{\phi_0 (x)\sqrt{V(x)}}{\phi_0
(x-i)\sqrt{V^*(x-i)}}\right)e^{-i\partial_x}V^*(x). \label{t1}
\end{eqnarray}
By virtue of (\ref{Cond-1}), the term in the bracket of the r.h.s.
of (\ref{t1}) is simply unity. Hence
\begin{equation}
\phi_0(x)
\sqrt{V(x)}\,e^{-i\partial_x}\sqrt{V^*(x)}\,\phi_0^{-1}(x)
=e^{-i\partial_x}V^*(x).
\end{equation}
Similarly, using (\ref{Cond-2}), the second term in $H$ transforms
as
\begin{eqnarray}
\phi_0(x)
\sqrt{V^*(x)}\,e^{i\partial_x}\sqrt{V(x)}\,\phi_0^{-1}(x)
=e^{i\partial_x}V(x).
\end{eqnarray}
Putting everything together, the result is
\begin{gather}
  L_{FP}=-e^{i\partial_x}V(x) -  e^{-i\partial_x}V^*(x)
  + V(x)+V^*(x).
  \label{FP1}
\end{gather}
This is the general form of FP operator corresponding to the
discrete Hamiltonian $H$ in (\ref{H}). One has $L_{FP}\phi_0^2=0$
as a consequence of $H\phi_0=0$.  Thus $\phi_0^2$ is the
stationary solution of the respective FP equation.

 We now discuss the limiting form of the FP equation when the
momentum $p=-i\partial_x$ is small.  We expand the operators
$e^{\pm i\partial_x}$ in (\ref{FP1}) up to the 2nd order in $p$.
The FP operator becomes
\begin{eqnarray}
L_{FP}=\frac{\partial}{\partial x} 2Im~V(x) +
\frac{\partial^2}{\partial x^2}2 Re~V(x).\label{n-rel}
\end{eqnarray}
Hence, in this limit, the deformed FP equation
(\ref{FP1}) does reduce to the usual FP equation (\ref{FPE}), with
$D^{(1)}=-2Im~V(x)$ and $D^{(2)}=2Re~V(x)$.

Let us illustrate this connection with a simple example discussed
in \cite{OS1}, namely, the discrete Schr\"odinger equation with
the Meixner-Pollaczek polynomials as eigenfunctions.  These
polynomials are 
deformation
of the Hermite polynomials.
The potential in this system is $V(x)=a+ix$ with $a$ real and
positive.  From the above discussion, the small momentum limit of
the FP equation is just the ordinary FP equation  with
$D^{(1)} = -2x$ and $D^{(2)}=2a$.  This latter system is none
other than the FP equation of the celebrated Ornstein-Uhlenbeck
process, which can be exactly solved by means of expansion in
terms of the Hermite polynomials \cite{FP}.

\medskip

{\bf 4.}~In \cite{OS1} four shape-invariant potentials \cite{shape}
in the
discrete quantum mechanics were presented, including the
Meixner-Pollaczek case discussed above. Shape invariance ensures
that the corresponding Schr\"odinger equations are exactly
solvable. Hence, the corresponding deformed FP equations are also
exactly solvable. For completeness, we list below the potentials
$V(x)$, the corresponding ground state wave functions $\phi_0(x)$,
and the eigenvalues $\lambda_n$ ($n=0,1,2,\ldots$). Needless to
say all the eigenfunctions are square integrable. All parameters
$a$, $b$, $c$ and $d$ are assumed to be real and positive.
We name the case by the
type of polynomials to which the polynomial part of $\phi_n$
belongs (see \cite{OS1} for the details of these polynomials).

\bigskip
\noindent
(i) the Meixner-Pollaczek case,\\
(one-parameter deformation of the Hermite polynomial)
\begin{gather}
  \qquad V(x)=a+ix, \nonumber\\
  \qquad\phi_0(x)\propto |\Gamma(a+ix)|, \nonumber\\
  \lambda_n=2n  \nonumber;
\end{gather}

\noindent
(ii) the continuous Hahn case,\\
(two-parameter deformation of the Hermite polynomial)
\begin{gather}
  \qquad V(x)=(a+ix)(b+ix), \nonumber\\
  \qquad\phi_0(x)\propto |\Gamma(a+ix)\Gamma(b+ix)|, \nonumber
  \\
  \lambda_n=n(n+2a+2b-1); \nonumber
\end{gather}
\noindent
(iii) the continuous dual Hahn case,\\
(two-parameter deformation of the Laguerre polynomial)
\begin{gather}
  \qquad V(x)=\frac{(a+ix)(b+ix)(c+ix)}{2ix(2ix+1)},
   \nonumber\\
  \qquad\phi_0(x)\propto
 \left|\frac{\Gamma(a+ix)\Gamma(b+ix)\Gamma(c+ix)}
  {\Gamma(2ix)}\right|, \nonumber\\
\lambda_n=n; \nonumber
\end{gather}

\noindent
(iv) the Wilson case, \\
(three-parameter deformation of the Laguerre polynomial)
\begin{gather}
  \qquad V(x)=\frac{(a+ix)(b+ix)(c+ix)(d+ix)}{2ix(2ix+1)},
   \nonumber\\
  \qquad
  \phi_0(x)\propto
\left|\frac{\Gamma(a+ix)\Gamma(b+ix)\Gamma(c+ix)\Gamma(d+ix)}
  {\Gamma(2ix)}\right|, \nonumber\\
\lambda_n=n(n+a+b+c+d-1). \nonumber
\end{gather}
For compactness, we have written the ground state wave function in
terms of the absolute value symbol.  For instance,
\begin{eqnarray}
|\Gamma(a+ix)| =\sqrt{\Gamma(a+ix)\Gamma(a-ix)}.
\end{eqnarray}

\medskip

{\bf 5.}~The potentials listed in the last section are exactly
solvable as they are shape-invariant. However, the construction of
the deformed FP operator can be carried over to the more general
potential
\begin{equation}
V(x)=\frac{\prod_j (a_j + ix)}{\prod_k (d_k + ix)}, \label{V-gen}
\end{equation}
where $a_j,~d_k$ are real and positive, and the degree of the numerator is at
least one greater than
  that of the denominator for square integrability.
The corresponding ground state wave function is
\begin{equation}
\phi_0 (x)\propto \sqrt{\frac{\prod_j \Gamma(a_j + ix)\Gamma(a_j -
ix)}{\prod_k \Gamma(d_k + ix)\Gamma(d_k - ix)}}.
\end{equation}
The potential $V(x)$ and the state $\phi_0 (x)$ satisfy the
relation (\ref{Cond}).  As discussed previously, this ensures that
the Hamiltonian $H$ with $V(x)$ in (\ref{V-gen}) is mapped into
$L_{FP}$ in (\ref{FP1}) with the similarity transformation
(\ref{FP}), and that the probability density $\phi_0^2$ is indeed
the zero mode of $L_{FP}$.

\medskip

{\bf 6.}~We turn now to a different type of shape-invariant
``discrete" quantum mechanical single particle systems with a
$q$-shift type kinetic term discussed in \cite{OS2}.
Eigenfunctions in this class of system are related to the
Askey-Wilson polynomials \cite{AW}.   We now derive the associated
FP operator for such systems.

Following \cite{OS2}, we use variables $\theta$, $x$ and $z$,
which are related as
\begin{equation}
  0\leq\theta\leq\pi,\quad x=\cos\theta,\quad z=e^{i\theta}.
\end{equation}
The dynamical variable is $\theta$ and the inner product is
$\langle f|g\rangle=\int_0^{\pi}d\theta f(\theta)^*g(\theta)$.
We denote $D\equiv z\frac{d}{dz}$. Then $q^D$ is a $q$-shift
operator, $q^Df(z)=f(qz)$.  We note here that
\begin{gather}
  \int_0^{\pi}d\theta
 =\int_{-1}^1\frac{dx}{\sqrt{1-x^2}},\nonumber\\
    -i\frac{d}{d\theta} =z\frac{d}{dz}=D,\quad
  f(z)^*=f^*(z^{-1}).
\end{gather}

For a function $V(z)$, which is a function of  a real constant $q$
($0<q<1$) and a set of real parameters, we define the following
Hamiltonian $H$,
\begin{eqnarray}
  H &\equiv &\sqrt{V(z)}\,q^{D}\!\sqrt{V^*(z^{-1})}
  +\sqrt{V^*(z^{-1})}\,q^{-D}\!\sqrt{V(z)}\nonumber\\
  &&-\left(V(z)+V^*(z^{-1})\right).
  \label{H-q}
\end{eqnarray}
The eigenvalue equation reads $H\phi_n=\lambda_n\phi_n$ with
eigenfunctions $\phi_n(z)$ and eigenvalues $\lambda_n$
($n=0,1,\ldots$) (we assume non-degeneracy
$\lambda_0<\lambda_1<\cdots$). The kinetic term causes a $q$-shift
in the variable $z$. This Hamiltonian is factorized, i.e.
$H=A^{\dagger}A$, with
\begin{eqnarray}
 A&=&
 q^{\frac{D}{2}}\sqrt{V^*(z^{-1})}
  -q^{-\frac{D}{2}}\sqrt{V(z)},\\
  A^{\dagger}&=&
 \sqrt{V(z)}\,q^{\frac{D}{2}}
  -\sqrt{V^*(z^{-1})}\,q^{-\frac{D}{2}}.
\end{eqnarray}

The ground state $\phi_0$ is the function annihilated by $A$:
\begin{equation}
  A\phi_0=0\quad(\Rightarrow H\phi_0=0,\ \lambda_0=0).
  \label{ground}
\end{equation}
Explicitly this equation reads
\begin{equation}
  \sqrt{V^*(q^{-\frac12}z^{-1})}\,\phi_0(q^{\frac12}z)
  =\sqrt{V(q^{-\frac12}z)}\,\phi_0(q^{-\frac12}z).
  \label{Cond-q}
\end{equation}
 Note that Eq.~(\ref{Cond-q}) implies
\begin{gather}
\sqrt{V^*(q^{-1}z^{-1})}~\phi_0(qz) =
\sqrt{V(z)}~\phi_0(z),\label{Cond-q1}\\
\sqrt{V^*(z^{-1})}~\phi_0(z) =
\sqrt{V(q^{-1}z)}~\phi_0(q^{-1}z).\label{Cond-q2}
\end{gather}
As with (\ref{Cond}), (\ref{Cond-q}) is the analogue of (\ref{W}).
The other eigenfunctions can be obtained in the form
\begin{equation}
  \phi_n(z)\propto p_n(z) \phi_0(z),
  \label{excited}
\end{equation}
where $p_n(z)$ is a Laurent polynomial in $z$ \cite{OS2}.

Using $H$, $\phi_0$, (\ref{Cond-q1}) and (\ref{Cond-q2}), the
similarity transformation (\ref{FP}) produces the FP operator
\begin{eqnarray}
  L_{FP}&=& - q^{-D}V(z)-
  q^{D}V^*(z^{-1})~\nonumber\\
  && + V(z)+V^*(z^{-1}).
  \label{L-q}
\end{eqnarray}
This is the general discrete $q$-deformed FP operator
corresponding to the Hamiltonian (\ref{H-q}).  Again, $L_{FP}$
annihilates $\phi_0^2$.

As an example, let us take $V(x)$ to be the one discussed in
\cite{OS2}:
\begin{equation}
  V(z)=\frac{(1-az)(1-bz)(1-cz)(1-dz)}{(1-z^2)(1-qz^2)}.
  \label{V-q}
\end{equation}
For simplicity we assume $-1<a,b,c,d<1$. Note that $V^*(z)=V(z)$.
The ground state is given by \cite{KS}
\begin{eqnarray}
\phi_0(z)&\propto &\left|
  \frac{(z^2;q)_{\infty}}{(az,bz,cz,dz;q)_{\infty}}\right|\\[3pt]
  &=&\sqrt{\frac{(z^2,z^{-2};q)_{\infty}}
  {(az,az^{-1},bz,bz^{-1},cz,cz^{-1},dz,dz^{-1};q)_{\infty}}},
\nonumber
\end{eqnarray}
where $(a_1,\cdots,a_m;q)_{\infty}=\prod_{j=1}^m
\prod_{n=0}^{\infty}(1-a_jq^n)$. Excited states have the form
(\ref{excited}) $\phi_n(z)\propto p_n(z) \phi_0(z)$, where
$p_n(z)$ is proportional to the Askey-Wilson polynomial
\cite{KS,AW}, which is a three-parameter deformation of the Jacobi polynomial.
The eigen-energies are
\begin{equation}
 \lambda_n
  =q^{-n}(1-q^n)(1-abcdq^{n-1}).
\end{equation}
Being shape-invariant, the Schr\"odinger equation of this system
is exactly solvable, and so is the corresponding FP equation.

\begin{acknowledgments}

This work is supported in part by the National Science Council
(NSC) of the Republic of China under Grant NSC 95-2112-M-032-012
(CLH), and in part by the Grant-in-Aid for Scientific Research
from the Ministry of Education, Culture, Sports, Science and
Technology under Grant No.18340061 and No. 16340040 (RS). We would
like to thank Koryu-kyokai (Japan) and  National Center for
Theoretical Sciences (Taipei) for support through the Japan-Taiwan
collaboration programs.  CLH and RS would like to thank the staff
and members of the Yukawa Institute for Theoretical Physics at the
Kyoto University and the National Taiwan University, respectively,
for their hospitality and financial support during their
respective visits.

\end{acknowledgments}

\end{document}